\documentclass{sig-alternate}
\graphicspath{ {Figures/} }
\usepackage{epstopdf}

\begin{document}
%

\title{Identifying Relevant Messages in a Twitter-based Citizen Channel for Natural Disaster Situations}
%
%
%
%
%

\numberofauthors{3} 
%
\author{
%
%
\alignauthor
Alfredo Cobo\\
       \affaddr{Pontificia Universidad Cat\'olica de Chile}\\
       \affaddr{Departamento de Ciencia de la Computaci\'on}\\
       \affaddr{Av. Vicu\~na Mackenna 4860, Macul}\\
       \affaddr{Santiago, Chile}\\
       \email{ajcobo@uc.cl}
\alignauthor 
Denis Parra\\
       \affaddr{Pontificia Universidad Cat\'olica de Chile}\\
       \affaddr{Departamento de Ciencia de la Computaci\'on}\\
       \affaddr{Av. Vicu\~na Mackenna 4860, Macul}\\
       \affaddr{Santiago, Chile}\\
       \email{dparra@ing.puc.cl}
\alignauthor
Jaime Nav\'on\\
       \affaddr{Pontificia Universidad Cat\'olica de Chile}\\
       \affaddr{Departamento de Ciencia de la Computaci\'on}\\
       \affaddr{Av. Vicu\~na Mackenna 4860, Macul}\\
       \affaddr{Santiago, Chile}\\
       \email{jnavon@ing.puc.cl}
}

\maketitle

\begin{abstract}
\label{chapter:abstract}
During recent years the online social networks (in particular Twitter) have become an important alternative information channel to traditional media during natural disasters, but the amount and diversity of messages poses the challenge of information overload to end users. The goal of our research is to develop an automatic classifier of tweets to feed a mobile application that reduces the difficulties that citizens face to get relevant information during natural disasters. In this paper, we present in detail the process to build a classifier that filters tweets relevant and non-relevant to an earthquake. By using a dataset from the Chilean earthquake of 2010, we first build and validate a ground truth, and then we contribute by presenting in detail the effect of class imbalance and dimensionality reduction over 5 classifiers. We show how the performance of these models is affected by these variables, providing important considerations at the moment of building these systems.
\end{abstract}



\section{Introduction}
\label{chapter:introduction}
In the minutes immediately after a catastrophe such as an earthquake or a tsunami, affected people experiment an urgent need for information of different kinds. First about the event itself, how big it was, where the epicenter was, or potential replicas. People also need to know about their relatives and friends, and the main source of information in the minutes after a quake has been traditionally the radio. There is however a lapse of time where the radio has no much information to communicate and they just broadcast anecdotal information about what is happening near the station or where some of the reporters happen to be at that moment \cite{Puente13}.
In the last years, people are turning to online social networks and in particular to Twitter to learn what is going on during and immediately after a catastrophic event.  This is especially true among youngsters who carry their smartphones at all times \cite{Valenzuela12}. Twitter has two big advantages as a news channel over the radio: first a very fast propagation speed \cite{Kwak10} and second it is bidirectional, that is, everyone can contribute with his own contents to the message stream.

Motivated by the potential impact of this technology in a country like Chile which suffers from frequent natural disasters and which population has adopted Internet through smartphones and online social networks at some of the fastest growing rates in Latin America \cite{Andrews13}, we have built an application with the purpose of serving as a citizen channel for disasters situations. Previous work have focused on identifying the credibility of tweet messages or in building tools for officials (government offices, response services, etc.) to help mitigating the effects of the events. However, to the best of our knowledge now such a tool is available for Spanish spoken audience and some important details on building the classification algorithms are not disclosed. In this paper, we provide details in the process of producing one of the key components of our application: an automatic classifier of tweets in Spanish language, which separates messages relevant from non-relevant to the disaster event. We contribute by providing details of the effect of number of latent dimensions and class imbalance in the performance of five different classifiers, which can be helpful to those building these type of methods based on machine learning techniques. 

\begin{figure}[ht]
\centering
\includegraphics[width=8cm]{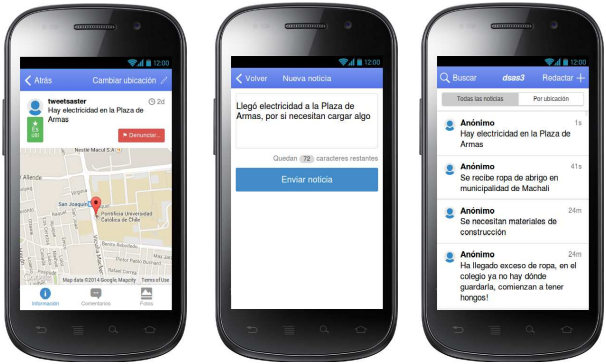}
\caption[Web application]{The mobile web application using the citizen channel.}
\label{fig:mobile}
\end{figure}

To train the classifier we used a stream of Twitter messages that was captured the minutes after the major Chilean earthquake of 2010.  To this end, the training set of messages were labeled as relevant or non relevant by human classifiers so this could be used as a "ground truth".
Our classifier is the most important piece of the citizen channel solution architecture that affected people can access through their mobile devices, to get relevant information and also to post new disaster related information that can be used by others.  Figure \ref{fig:mobile} provides a few snapshots from the mobile web application in action \cite{Molina15}.

The rest of the paper is organized as follows.  In the next section we provide a review of the most relevant literature and related work.  In Chapter \ref{chapter:methodology} we explain the methodology we used to built the automatic classifier.  Chapter \ref{chapter:results} presents the results we got when we trained our classifier to close in Chapter \ref{chapter:conclusions} with conclusions and future work.

\section{Related work}
\label{chapter:relatedwork}
In order to present the related work we divided them into several groups. First manual classification research and post processing are presented. Other feature approaches and analysis are shown. Then several tools for disaster management are reviewed, finishing with our particular goals.

\textbf{Manual classification.} There have been many attempts to capture and process the twitter messages generated in situations of natural disasters. The first attempts were simple manual classification. Vieweg et al. \cite{Vieweg10} manually classified situational messages about Oklahoma Grassfires of April 2009 and the Red River Floods that occurred in March and April 2009. Imran et al.\cite{Imran13nugget} did the same process for the Joplin tornado of 2011 but he used crowdsourcing services afterwards to perform automatic classification using machine learning techniques. Nevertheless we are not aware of any real time Spanish language automatic classification attempt needed to feed a citizen information channel for natural disaster events. 

\textbf{Post processing.} Regarding post processing of the messages there is also relevant work. Castillo et al. \cite{Castillo11} assessed the credibility of the messages, while Mendoza et al. \cite{Mendoza10} classified dissemination of false rumors and confirmed news of the Chilean 2010 earthquake. We addressed relevance of messages according to a certain criteria, using post processing similar to these works.

\textbf{Feature generation approaches.}
There have also been attempts to improve the performance of the algorithms by generating new features. For example Gimplel et al. \cite{Gimpel11} used part of speech recognition in English while Kouloumpis et al. \cite{Kouloumpis11} and Liu et al. \cite{Liu12} used several tools such as sentiment analysis to add features to the training set. Another type of analysis can be made over a generated network graph. Wu et al.\cite{Wu11} examined the information generated and consumed by Twitter users, resulting in distinguishable groups and high concentration. Lee et al. \cite{Lee14} studied the likelihood of a user to make a retweet to spread information. We obtained similar features, based on the ones available in the 2010 earthquake and considering real time restrictions. Features over the network were not considered in particular due to this restriction. 

\textbf{Tools for disaster management.}
There have been several attempts in constructing frameworks to deal with the information overload produced by twitter messages \cite{Hiltz13}. Power et al. \cite{Power13} characterized tweets as a fast source of information for situation awareness. Caragea et al. \cite{Caragea11} build a framework to aid NGOs and first responders to record and classify and aggregate data from the Haiti 2010 earthquake. Abel et al. \cite{Abel12} made a tool to explore information from Twitter and other web streams. Middleton et al. \cite{Middleton14} developed a decision support system to give awareness in earthquake and tsunami events. Morstatter et al. \cite{Morstatter13xplorer} created a system to gain knowledge and visualize events. Recently research has been done in crowdsourced tagging, so the algorithms can be repeatedly trained over time. Imran et al. \cite{Imran14} proposed a framework to actively tag messages during an event.

Almost all these frameworks were designed to help official agencies and focus their efforts to suit their needs. However they tend to forget that citizens are also in need of situational awareness, to take informed decisions. In our research the product has been targeted to them, affecting our assumptions and decisions.
These frameworks are generally designed to help the official agencies and tend to forget that the citizen are also in need of situational awareness. In our research the final product would be targeted to them, affection our assumptions and decisions.

Previous research efforts served as a guideline for our work. The main differences were context and focus. A Spanish channel for earthquake situations guided to Chileans and focused on the citizens. Gathering relevant messages and delivering them in real time to them. The challenge was to know if a classifier with could be made addressing these characteristics.


\section{Context}
\label{chapter:context}
Twitter can then have an important role in informing citizens during disasters specially in countries frequently suffering from these events and where internet, smartphones and onine social networks have shown a quick penetration. Chile is one of such countries. It suffered from a major earthquake (8.8 Richter) not far ago in 2010 and new one (8.2 Richter) in 2014, and the largest ever registered earthquake took place in Chilean territory in 1960. In Chile there has been a very fast penetration of mobile devices and a large segment of the population owns a smartphone \cite{Ureta08}. 
Chilean authorities have taken notes of both the rising popularity of Twitter and the ubiquity of smartphones and they have open Twitter accounts to inform the citizens. For instance, ONEMI (National Office for Emergencies), SHOA (Army Hydrographic Service) and others tweet every time an important event occurs. However, there are still some barriers. One problem to adopt Twitter as a main source of emergency news is that for an important segment of the population it is complicated. For a senior citizen, to create an account and then to follow the relevant sources, not to mention the possibility to write his own messages can be nearly impossible. To face this problem we built a friendly web application that lowers the technology barriers.
But the main problem of Twitter was the noisy nature of the channel, that can produce information overload \cite{Hiltz13}. Together with those messages from ONEMI and SHOA the user will be getting many non-relevant messages in his timeline that could hide the important ones.
Using machine learning algorithms, we designed and build an automatic classifier which was able to classify a message as "relevant" or "non-relevant" where relevants were the ones that contained some information relative to an earthquake event.

\section{Methodology}
\label{chapter:methodology}
Our aim was training a model that could be able to predict relevant messages, taking a supervised learning approach. To do this we first created a ground truth due to the lack of labeled earthquake Spanish data available. In the next subsections, creation and validation of a suitable dataset are shown, followed by feature selection and the evaluation procedure for the final model.


\subsection{Building the ground truth}

The tweets used to build our ground truth were obtained from a known earthquake dataset \cite{Mendoza10}, which were posted before and after the critical event (2010-02-27 03:34:08). They started at midnight of February 27th and ended at midnight of March 2nd. These data were not labeled beforehand, and the relevance of messages was not explicit. Therefore, we built a ground truth performing manual labeling so it could be used to train supervised learning classifiers.

Due to limited resources and time constraints we gathered a subset of the whole dataset, so we could have a fine control over each message to be labeled. It was important also to have control over the people that were going to classify, because of the Chilean context and local terms that appeared in the data.

A subset of 5000 tweets was initially obtained using systematic sampling (a similar number of messages per each day), to have a more homogeneous set over time. After this, we removed messages which were not written in Spanish by using the language processing tools textcat\footnote{http://CRAN.R-project.org/package=textcat} and tm\footnote{http://CRAN.R-project.org/package=tm} packages, followed by a manual inspection of every message. 
Subsequently, tweets with too similar phrases were removed using 10\% Lavenshtein distance as minimum tolerance. Afterwards a manual review was done, to ensure low redundancy. All of these steps resulted in a final dataset of 2187 messages: 524 tweets for day one, 529 for day two, 618 for day three and 516 for day four.

Once the base dataset was defined, we provided the labelers (three people per each tweet) a known criteria to identify each message \cite{Imran13nugget}. The goal was to classify the tweets into one of two classes, either ``relevant'' to the earthquake situation if the message belonged to any of these categories, or ``not relevant'' if the subject deemed it not related to the disaster. Then, the tweets were deemed relevant if any of the following criteria was met:

\begin{itemize}
  \item \textbf{Caution and advice.} The message conveys/reports information about some warning or a piece of advice about a possible hazard of an incident. 
  \item \textbf{Casualties and damage.} The message reports the information about casualties or damage done by an incident.
  \item \textbf{People missing, found, or seen.} The message reports about the missing or found person affected by an incident or seen a celebrity visit on ground zero. 
  \item \textbf{Information source.} The message conveys/contains some information sources like photo, footage, video, or mentions other sources like TV, radio related to an incident.
\end{itemize}

Using these categories, 6 people classified the dataset, dividing them in groups to produce 3 labels for each message. 

\subsection{Validation of the ground truth}

In order to set a unique label for each tweet, we used majority vote as criteria, so if at least two users agreed on the label of the tweet, we used that label as gold standard. To enhance the validity of our ground truth, we performed an additional validation step analyzing people's agreement. 

The raw agreement, calculated as the proportion of agreements divided by all the possible cases of agreement, was 74.2\%. This can be considered a reasonable agreement between all raters. We assessed the reliability of agreement among the people labeling the tweets using Fleiss' kappa which, unlike Cohen's kappa, it allows to measure the agreement between more than two raters. We obtained a significant substantial agreement of $\kappa=0.645$, $p<.001$, as explained by Landis and Koch \cite{Landis77}. We also calculated the raters agreement by intraclass correlation using the ICC(2,1) model, which resulted in a $IIC=0.646$, with $F(2198,3585)=6.54$, $p<.001$. This is considered a moderate agreement \cite{McGraw96}. 

\subsection{Model features}
After building and validating the ground truth, the next step was constructing the classifier. For that matter we explain the feature selection, dimensionality reduction and class imbalance problems.

\textbf{Feature set:}
The dataset had mainly two groups of selectable features: user-based and content-based. From the user we extracted the number of followers and friends, which are directly usable in the model. From the content we performed text preprocessing, including tokenization and Spanish snowball stemming. From the corpora we used hashtags, words and user mentions, removing everything else. 
The number of resulting features were 4766 using a tfidf vectorizer for the filtered content, considering a minimum frequency value of one word.


\begin{table*}[htb!]
\centering
  \begin{tabular}{ |c|c|c|c|c|c||c|c| }
    \hline
    Model & Precision & Recall & F1 score & Accuracy & AUC & Dimensions & Noise proportion \\
    \hline
    Baseline & 0.625 & 0.545 & 0.53 & 0.5 & 0.568 & - & 0 \\
    Bernoulli NB & \textbf{0.831} & 0.226 & 0.355 & 0.594 & 0.605 & 2000 & 0.0\\
    Logistic Regression & 0.827 & 0.641 & 0.722 & 0.756 & 0.834 & 1000 & 0.6\\
    Linear SVM & 0.687 & \textbf{0.677} & 0.682 & 0.687 & 0.719 & 1000 & 0.6\\
    Random Forest & 0.807& 0.673 & \textbf{0.734} & \textbf{0.758} & \textbf{0.844} & 1000 & 0.8\\
    \hline
  \end{tabular}
\caption[Scores of the best classifiers]{The best scores for each classifier. (For every score the best is marked in bold)}
\label{tab:results}
\vspace{-4mm}
\end{table*}

\textbf{Dimensionality reduction:} Previous works have used dimensionality reduction techniques to reduce the number of features of classifiers and to boost their performance. Newman et al. \cite{Newman07} and Biro et al. \cite{Biro08} increased their classifier performance using dimensionality reduction techniques over their baselines. Motivated by these works, we used 
Latent Dirichlet Allocation (LDA) to reduce the number of features. LDA was chosen over latent semantic indexing (LSI) because it could handle unseen documents giving a prediction when the words were not previously observed by the model \cite{Wei06}. 


\textbf{Class imbalance:} 
As previous works mention, adding new data as noise and balancing the classes can improve performance. Wang et al. presented the class imbalance as a problem that can reduce performance \cite{Wang12}. To address this issue we used the boundary SMOTE algorithm \cite{Han05} to over sample the relevant messages so the bigger datasets could be balanced. This was done before each round of training. 

In order to add the required noise (i.e., tweets not relevant to earthquakes) we gathered another set of tweets from Twitter streaming API, connecting a geographic localized query to the service for about 5 months, from 16/05/2014 to 27/10/2014. This query drew a rectangle over Chile, so every tweet in this dataset was from or nearby this country. Afterwards messages that were not recognized as Spanish by Twitter were removed. Additionally seismic activity related tweets were filtered from events starting at magnitude 4 Richter. The messages that were 20 minutes before until 2 hours after each event were removed. Systematic sampling was used to extract the messages from this filtered dataset in order to add them as noise to the ground truth before each training phase. The proportion of not relevant messages were added as a 20, 40, 60 and 80 percent of the ground truth length.

\subsection{Algorithms and evaluation procedure} 

To choose the best performing algorithm in our experiments, we compared four well-known classifiers (logistic regression, random forests, SVM and Bernoulli Naive Bayes). These algorithms are supervised, hence, in the training phase a  5-fold cross-validation procedure was used to tune parameters of these models. In addition to the aforementioned models, a baseline was defined using the presence of the word ``terremoto'' (``earthquake'' in Spanish). Thus, whenever this word showed up, the document was marked as relevant, performing better than random guessing. For all these models, the evaluation metrics were precision, recall, accuracy, F-score \cite{manning2008introduction} and AUC \cite{fawcett2004roc} as in previous works \cite{Wen2014,Castillo11}. Although we considered recall the most relevant criteria, since in our context we wanted to collect most of the relevant messages, we did not want to sacrifice too much precision so F-score and AUC were considered our main evaluation metrics.

\section{Results}
\label{chapter:results}

\begin{figure*}[htb!]
\begin{tabular}{cc}
 \includegraphics[width=83mm]{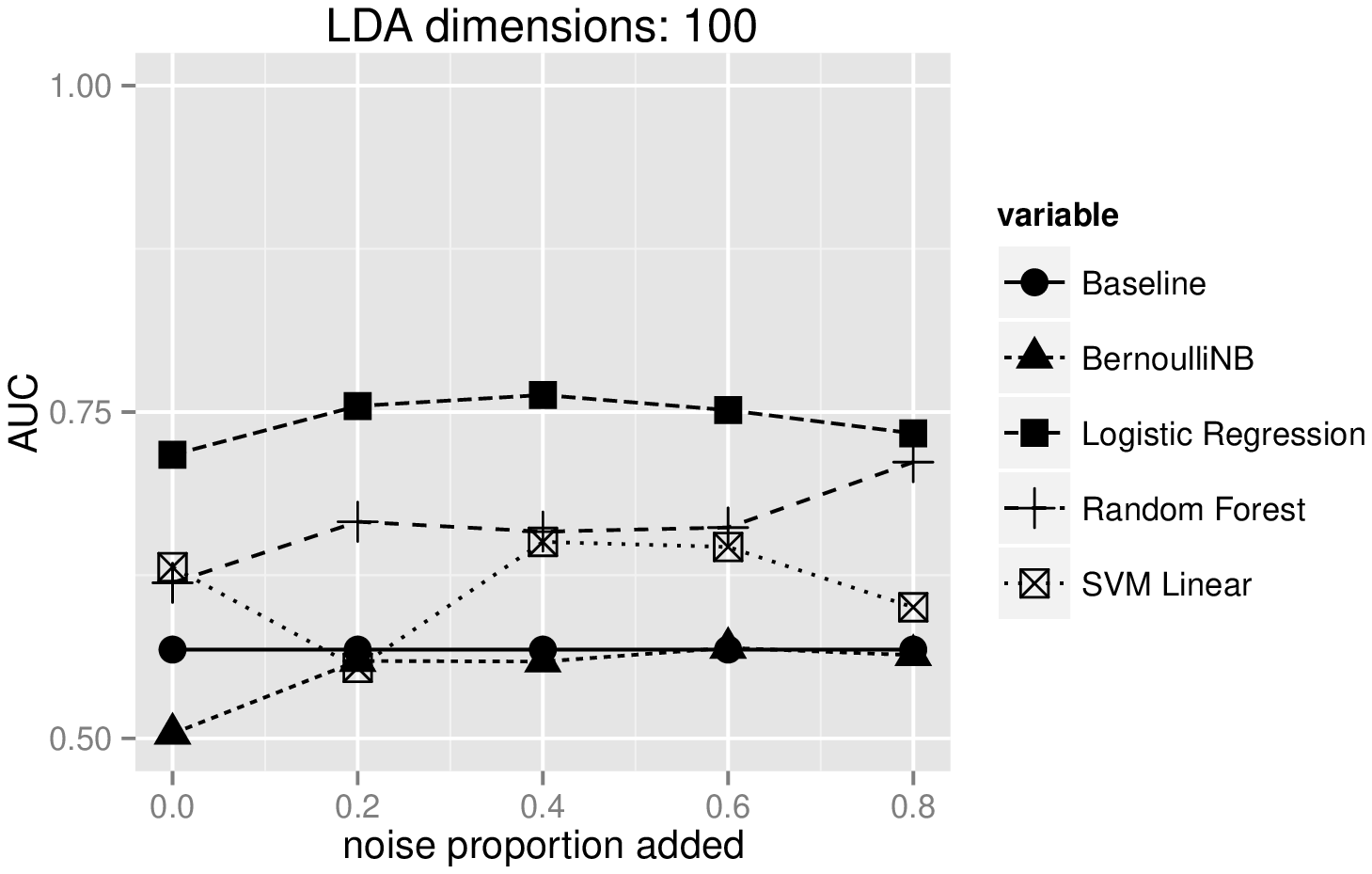} &  \includegraphics[width=83mm]{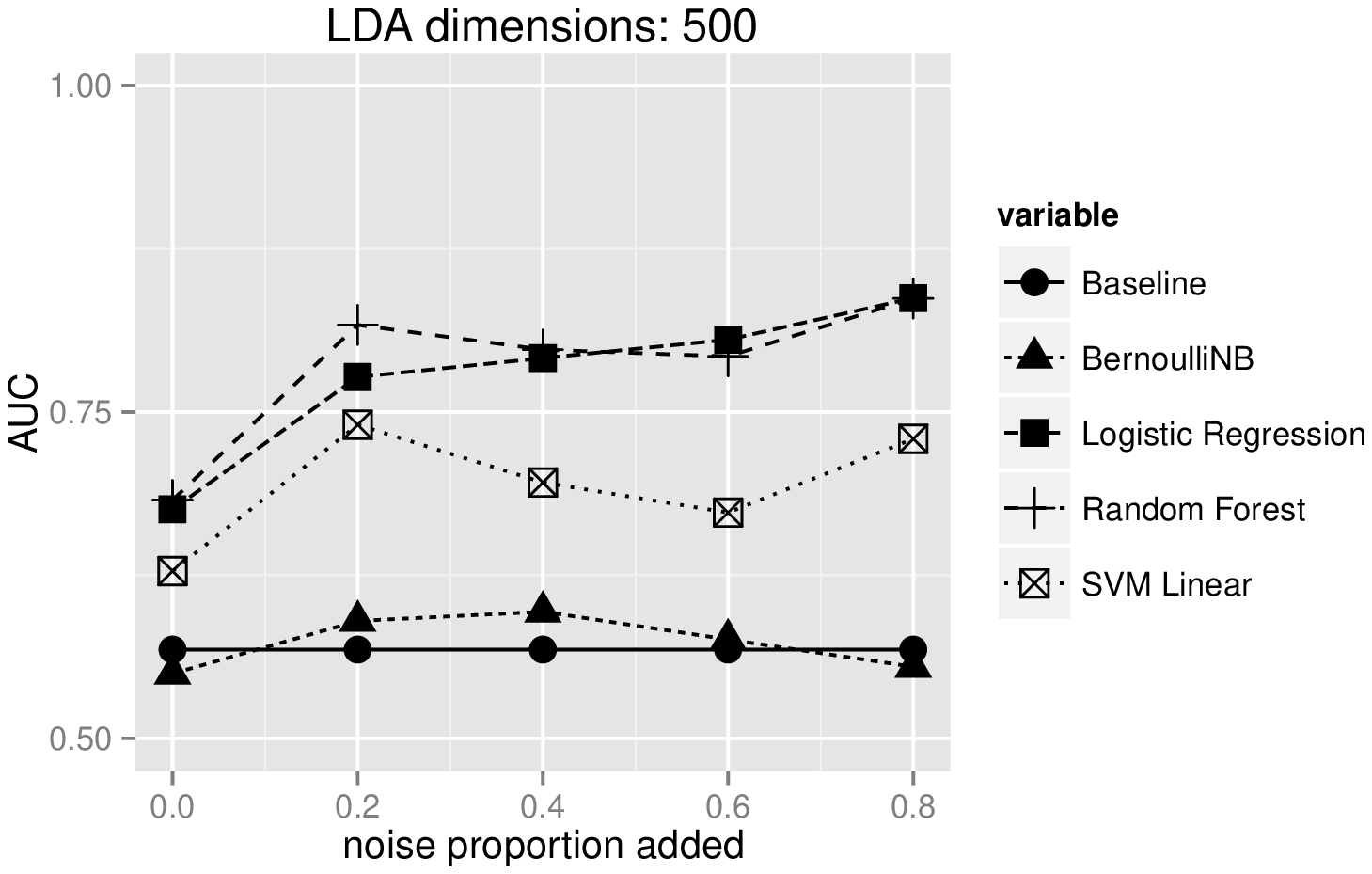} \\
 \includegraphics[width=83mm]{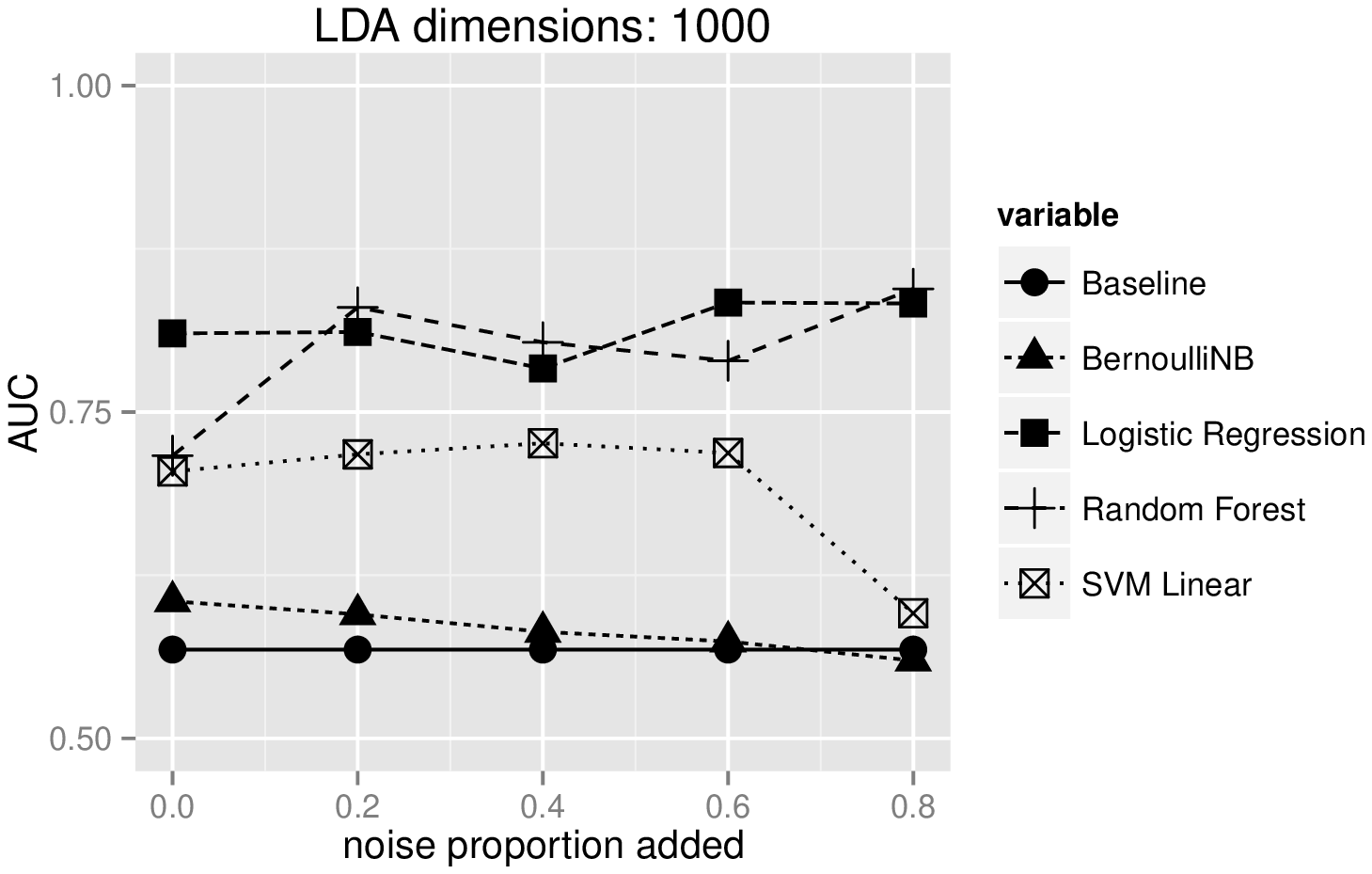} &  \includegraphics[width=83mm]{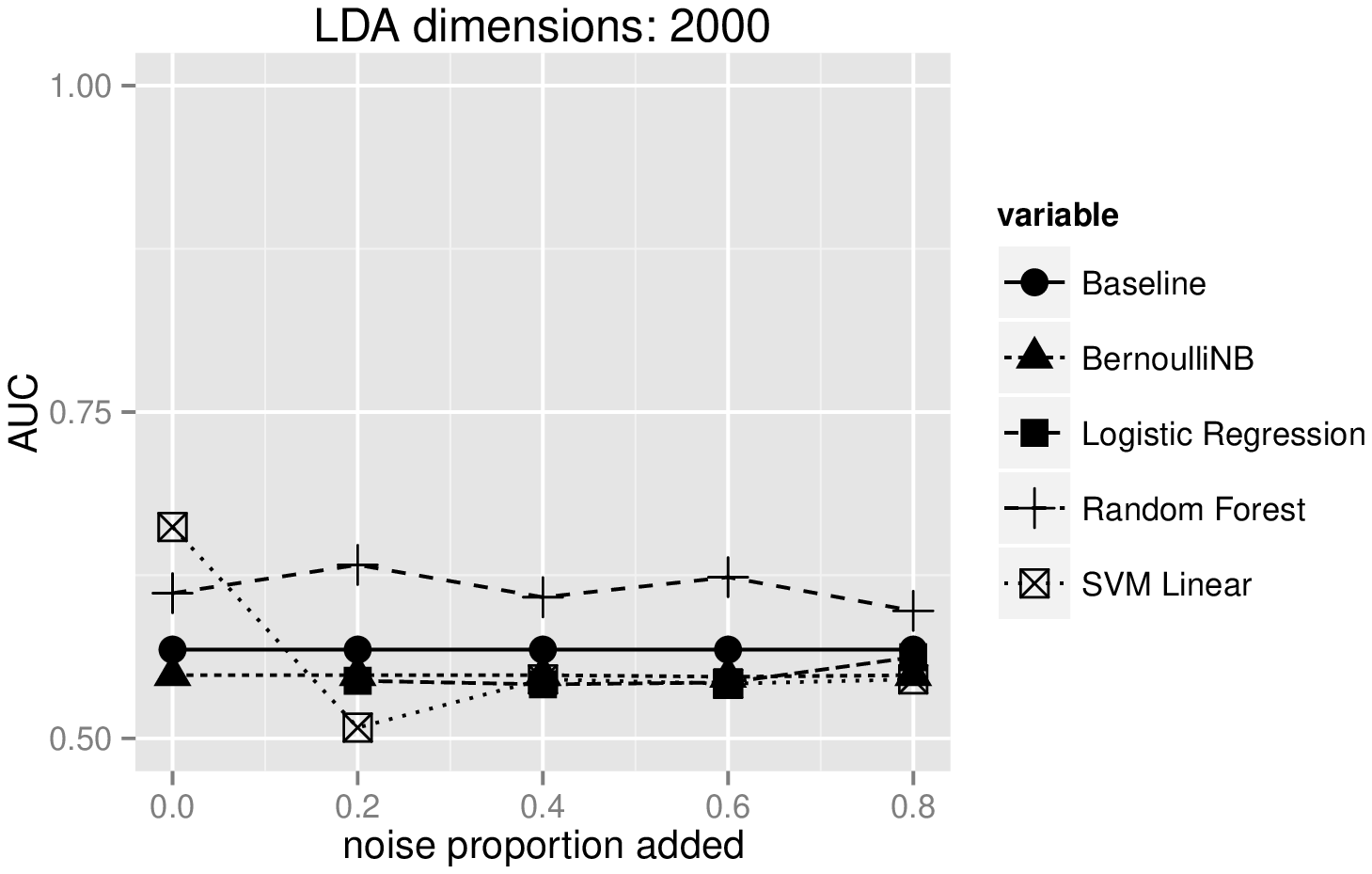} \\
\end{tabular}
\caption[Variation of AUC scores over LDA dimensions]{ Variation of AUC scores when latent dimensions are set to different values.}
\label{fig:LDAAUC}
\vspace{-6mm}

\end{figure*}

The results shown at Table \ref{tab:results} indicate that the best recall score was given by the linear SVM model with 0.677. However, random forests performed more consistently over several metrics that evaluate precision and recall. We preferred this model considering that its recall was slightly below the best one (0.673 versus 0.677), but outperformed it clearly in precision (0.807 versus 0.687) and as a consequence in other scores, such as F1 score where 0.734 improved over 0.682 of the SVM. Furthermore, the random forests showed the best adaptation of the whole set of models, performing better under more noisy data and at different number of latent dimensions, as could be observed in Figure \ref{fig:LDAAUC}. The four plots in Figure \ref{fig:LDAAUC}  show the behaviour of our models under different amount of LDA dimensions (100, 500, 1000 and 2000) and with different proportions of noisy data in the x-axis (from 0 to 0.8). Random forests, unlike the other models, is always able to increase or maintain its performance when the proportion of non-relevant tweets (noise) increases. Logistic regression shows a similar trend, even outperforming random forests when the number of LDA dimensions is small (100), but decreasing considerably when the dimensions increase to 2000. 


\textbf{Dimensionality reduction:} Figure \ref{fig:LDAAUC} shows that when the latent dimensions were among 500 and 1000, all the algorithms showed their best performance. However, with 2000 dimensions random forests tended to perform better than the rest, in a context where every algorithm performed poorly. This led us to think that some of the trained models presented a high variance problem, as explained by Amatriain \cite{Amatriain13}, who suggested that models that present high variance can be benefited from fewer features and more data. The effect of fewer features (dimensions) and more training samples had a positive impact on the scores, because the models were too complex for the amount of initial data (without noise) we had.

\textbf{Class imbalance:} When more samples of non-relevant tweets were added to the training dataset we observed the same tendency, as they increased the performances got better, as observed in Figure \ref{fig:LDAAUC}. In the case of 100 dimensions the algorithms did not benefit from it. However, between 500 and 1000 dimensions, logistic regression and random forests had a significant boost in performance. In fact these two models had their highest AUC scores in this region of parameters. 

In summary, improvement of model performance was more affected by the number of latent dimensions than by the additional training data provided, but still the top performing models (random forests and logistic regression) were improved by leveraging additional non-relevant training messages. Hence, we suggest including  these two factors for learning better models and for generalization of this classification problem.


\section{Conclusions}
\label{chapter:conclusions}
As the new generations take over and technology makes possible for anyone to own a sophisticated mobile device, Twitter and social networks will be used more and more to get fast information about special events. A natural disaster event is no exception and recent experience in Chile demonstrated the important role of social networks.
Our goal was to leverage the main advantages of Twitter to produce a citizen to citizen information channel which architecture we described in this paper. A key component of this architecture was an automated classifier that can filter the huge and noisy flow of twitter messages, discarding all messages that were not related to the event. This channel was used to feed a mobile web application that the citizen could use at the time of the event.
The building of the classifier involved many challenges including the definition of a reliable and validated ground truth and the selection of an appropriate algorithm in the context that the classifier was going to be used.
After analyzing and comparing several classifiers we finally could get one that performed remarkably well for the purposes of our citizen channel. The selected model was a random forest that had 0.807 precision, 0.673 recall and 0.844 AUC, outperforming our baseline and all other classifiers evaluated. Having this result allowed us to make a big step toward the implementation of our system.
Moreover, we identified the relevance of the number of features and amount of data in the training of the models. It was important to train considering these two subjects to have better performances overall, specially the ones that had high variance on our base dataset.
The final evaluation for this model will be conducted when the architecture is in full operation and more citizens use the application during a forthcoming seismic event.


\section{Acknowledgements}
\label{chapter:acknowledgements}
This work has been partially supported by the National Research Center for Integrated Natural Disaster Management CONICYT/FONDAP/15110017.

%
\bibliographystyle{abbrv}
\bibliography{bibliography} 
\end{document}